\documentclass[aps,prl,twocolumn,showpacs,floatfix]{revtex4}
\usepackage{graphicx}
\usepackage[colorlinks=false,citecolor=blue]{hyperref}
\begin{document}

\title{Emergence of turbulence in an oscillating Bose-Einstein condensate}

\author{E. A. L. Henn$^{1}$}\email{ehenn@ifsc.usp.br}
\author{J. A. Seman$^{1}$, G. Roati$^{2}$, K. M. F. Magalh\~{a}es$^{1}$ and V. S. Bagnato$^{1}$}
\affiliation{$^{1}$Instituto de F\'{\i}sica de S\~ao Carlos, Universidade de S\~ao Paulo, Caixa Postal 369, 13560-970 S\~ao Carlos, SP, Brazil\\$^{2}$LENS and Dipartimento di Fisica, Universita di Firenze, and INFM-CNR, Via Nello Carrara 1, 50019 Sesto Fiorentino, Italy}
\date{\today}

\begin{abstract}

We report on the experimental observation of vortex tangles in an atomic BEC of $^{87}$Rb atoms when an external oscillatory perturbation is introduced in the trap. The vortex tangle configuration is a signature of the presence of a turbulent regime in the cloud. We also show that this turbulent cloud has suppression of the aspect ratio inversion typically observed in quantum degenerate bosonic gases during free expansion. Instead, the cloud expands keeping the ratio between their axis constant. Turbulence in atomic superfluids may constitute an alternative system to investigate decay mechanisms as well as to test fundamental theoretical aspects in this field.
 
\end{abstract}


\maketitle

\newcommand{\be}{\begin{equation}}
\newcommand{\ee}{\end{equation}}
\newcommand{\bea}{\begin{eqnarray}}
\newcommand{\eea}{\end{eqnarray}}
\newcommand{\bi}{\bibitem}
\newcommand{\la}{\langle}
\newcommand{\ra}{\rangle}
\newcommand{\ua}{\uparrow}
\newcommand{\da}{\downarrow}
\renewcommand{\r}{({\bf r})}
\newcommand{\rp}{({\bf r'})}
\newcommand{\rpp}{({\bf r''})}

Turbulence in classical fluids has always been an intriguing issue in physics and understanding and controlling turbulence represents one of the main challenges in physics \cite{Landau}. Nevertheless, classical turbulence is a concept difficult to define and there are no fully consistent theories to describe it \cite{Drosdoff}. A turbulent flow is a highly unsteady flow characterized by the presence of vorticity, diffusion and energy dissipation processes. In a turbulent fluid, the fluctuation of pressure and velocity occur in a very broad range of length and time scales, making its mathematical description very complex \cite{Peric}. 

Turbulence is also possible in superfluids, such as superfluid Helium \cite{Donely} and superfluid $^3$He-B \cite{3b}. Turbulence in superfluids or Quantum Turbulence (QT) is characterized by the appearance of quantized vortices distributed in a tangled way \cite{QT, QT2, tsubotaturb}, not forming regular lattices.  This subject has been extensively studied in superfluid Helium both experimentally and theoretically since it has been discovered about 50 years ago \cite{Hall}.
Until recently, turbulence in He-superfluid could only be observed by indirect methods. Oscillating wires or wire grids \cite{Hemethod} were the most commonly employed techniques. Recently, there was the first direct observation of QT in Helium \cite{hydrogen}. That work opened up new frontiers on the study of QT in superfluid Helium \cite{newH}. In the theoretical branch, QT in Helium has been satisfactorily modeled by the vortex filament model \cite{vfm}, though some features such as vortex reconnections must be introduced artificially.  

Nevertheless, the achievement of BEC in trapped atomic gases \cite{bec} and the subsequent observation of quantized vortices in these samples \cite{vortex} opened up the possibility to study turbulence in a more controlled fashion, since one can control the main characteristics (interaction, atomic density, number of atoms, trapping configuration) of the atomic sample. Additionally, the possibility of expanding the quantum cloud by releasing the atoms from the trap, makes possible the direct observation of the vortices by optical means. There are today many open questions related to QT, including mechanisms of decay \cite{QT2} which may find important hints on the direct observation of QT in a trapped atomic superfluid. In fact, the study of QT in atomic superfluids may shine light on the turbulence characteristics that are universal, due to the superfluidity as a macroscopic quantum state, and reveal which characteristics depend on the specific nature of the superfluid. Specifically, in atomic superfluids, the ratio of the average vortex spacing to the vortex core size is expected to be much smaller than in Helium, leading to different modes of turbulence decay. Several extensive reviews \cite{vfm, review2, review3} summarize the present status and challenges to be faced in the research field of QT. 

Shortly after trapped BECs were first realized, the observation of vortex lattice structures, as well as the crystallization dynamics of these structures were reported \cite{abo}. These dynamics have been successfully confirmed quantitatively by the numerical simulation of the Gross-Pitaevskii equation \cite{vortexsimul}. However, in the experimental research of trapped BECs, turbulence has not yet been investigated. Recently, this topic has attracted theoretical interest \cite{svustunov, tsubotaturb,parker} and numerical simulations show that turbulence is probably present as a step in the formation dynamics of a BEC with the presence of spontaneously formed vortices \cite{Weiler}. Nevertheless, no experimental evidence of this regime has been specifically presented so far. In this Letter, we report on the experimental observation of QT in a magnetically trapped BEC of $^{87}$Rb atoms evidenced by the observation of tangled vortices in the quantum sample, an intrinsic characteristic of QT, as well as a change in the hydrodynamic behavior of the atomic cloud. 

\begin{figure}
\centering
\includegraphics[scale=0.3]{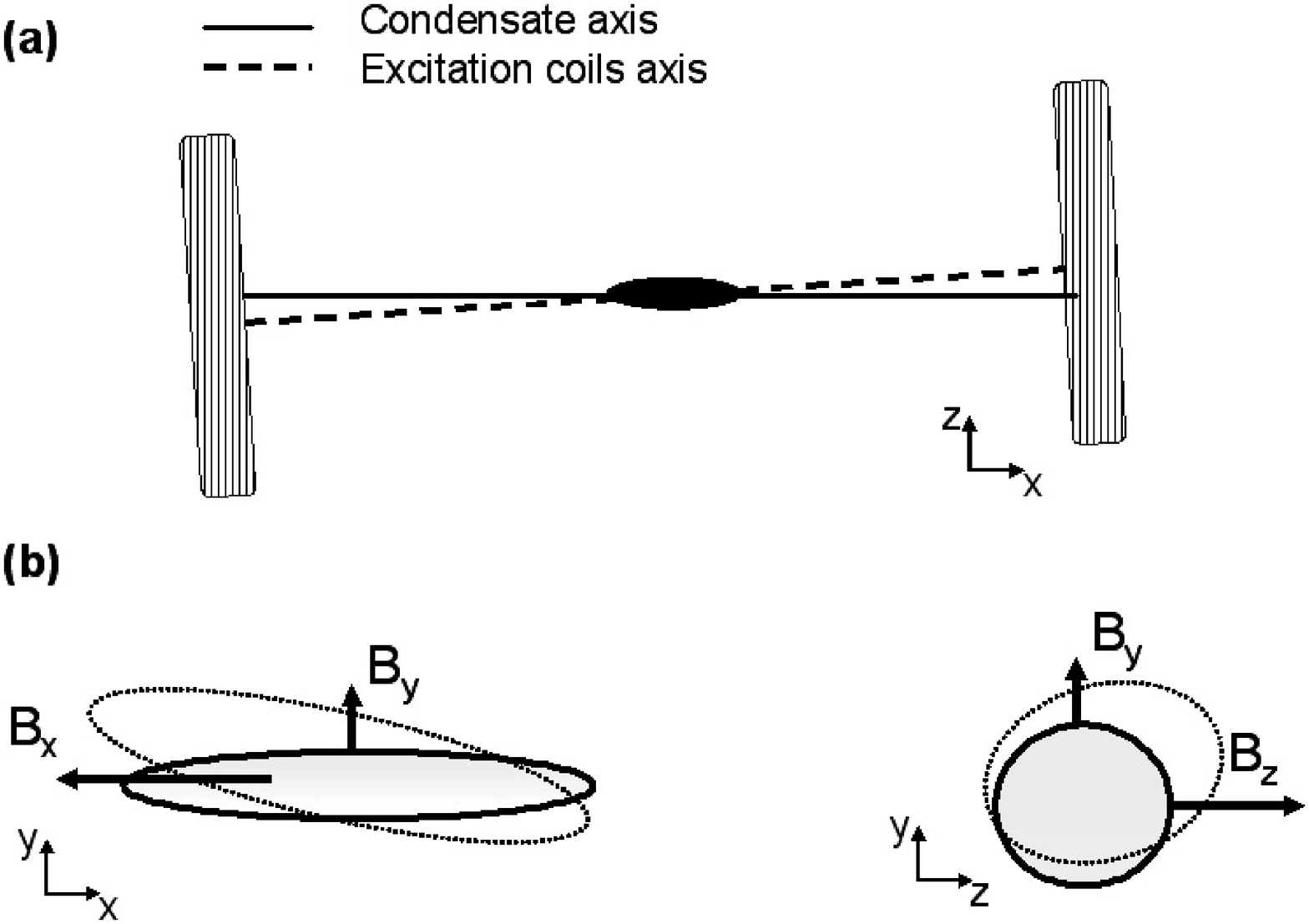}
\caption{(a) Experimental configuration to nucleate vortices in the BEC, showing the anti-Helmholtz coils slightly misaligned from the condensate long axis and (b) pictorial view of the expected movements/deformations suffered by the condensate cloud.}
\label{1}
\end{figure}

The experimental sequence to produce the BEC as well as the procedure to generate vortices in the condensate are described in details in Refs.\cite{BJP,vortexform}. In brief, we produce a BEC of $^{87}$Rb containing $\left(1-2\right)\times10^5$ atoms with a small thermal fraction in a cigar-shaped magnetic trap with frequencies given by $\omega_r=2\pi\times210Hz$ and $\omega_x=2\pi\times23Hz$. Once the condensate is obtained and still held in the trap, an oscillating magnetic field is applied. This field is produced by a pair of anti-Helmholtz coils placed with their axis closely, though not exactly, parallel to the weak axis of the trap (Fig.\ref{1}a). Since the coils are not perfectly aligned to the condensate axis, the external field in the region of the trap has components parallel to each trap eingenaxis. Additionally, the components along the two equal directions that give the radial symmetry of the trap are slightly different, breaking it up as well as the symmetry of oscillation of the condensate. This oscillatory scheme was developed to investigate coherent mode excitations in Bose-condensates \cite{Yuka}. Such excitations are also believed to produce vortices.

A pictorial view of the field acting on the condensate is shown in Fig.\ref{1}b. A combination of shape modification, displacement of the minimum and rotation is coupled to the atoms. The quadrupolar field is applied for a period ranging from 20 to 60 ms, oscillating sinusoidally in time with a frequency of 200Hz. This extra magnetic field has an offset so that in a period of oscillation, it goes from zero to a maximum value and back to zero, never inverting its direction. The maximum value reached by the field gradient characterizes the strength of the excitation and has been varied from zero to 190 mG/cm in the axis of the anti-Helmholtz coils. After the end of the oscillation stage, the atoms are left trapped for an extra 20 ms before being released and observed in free expansion by a standard absorption imaging technique. For small amplitudes of the oscillating field as well as short excitation periods we observe dipolar modes, quadrupolar modes and scissors modes \cite{scissor, scissor2} of the BEC but no appearance of vortices. Increasing both parameters we start to see vortices that grow in number with amplitude and/or duration of excitation. These regimes are well described in a previous report \cite{vortexform}.

The effective mechanism that nucleates the vortices is still under theoretical investigation \cite{mm}. We believe that a possible mechanism of vortex formation are the so-called Kelvin-Helmholtz instabilities \cite{KH}, a well-known phenomena in the scope of fluid interface theory and experiments. These instabilities occur at the interface of two fluids with a relative motion and give rise to the formation of vortices. This phenomenon has been theoretically investigated \cite{KH theory} in normal fluid-superfluid mixtures and directly observed in mixtures of A and B phases of superfluid liquid Helium \cite{KH superfluid}. The fact that this mechanism may be taking place is supported by previously theoretical and experimental observations \cite{parker, vortexform, scissor, scissor2}. Ref.\cite{parker} shows that a stirred condensate ejects energetic atoms to form an outer cloud during the quadrupolar mode excitation.  Additionally, Refs. \cite{scissor, scissor2} show that condensed, superfluid clouds and normal clouds have different frequencies of oscillation, leading to relative motion between both components of the fluid. Finally, in Ref.\cite{vortexform}, where we describe the vortex nucleation protocol, we directly observe the quadrupolar modes as well as vortex structures in the interface of the dense core that represents the condensed cloud and the outer cloud (Fig.7 of Ref.\cite{vortexform}). We thus believe that this is the effective mechanism contributing to nucleate the vortices and eventually lead to the turbulent state in the oscillating excitation.

\begin{figure}
\centering
 \includegraphics[scale=0.4]{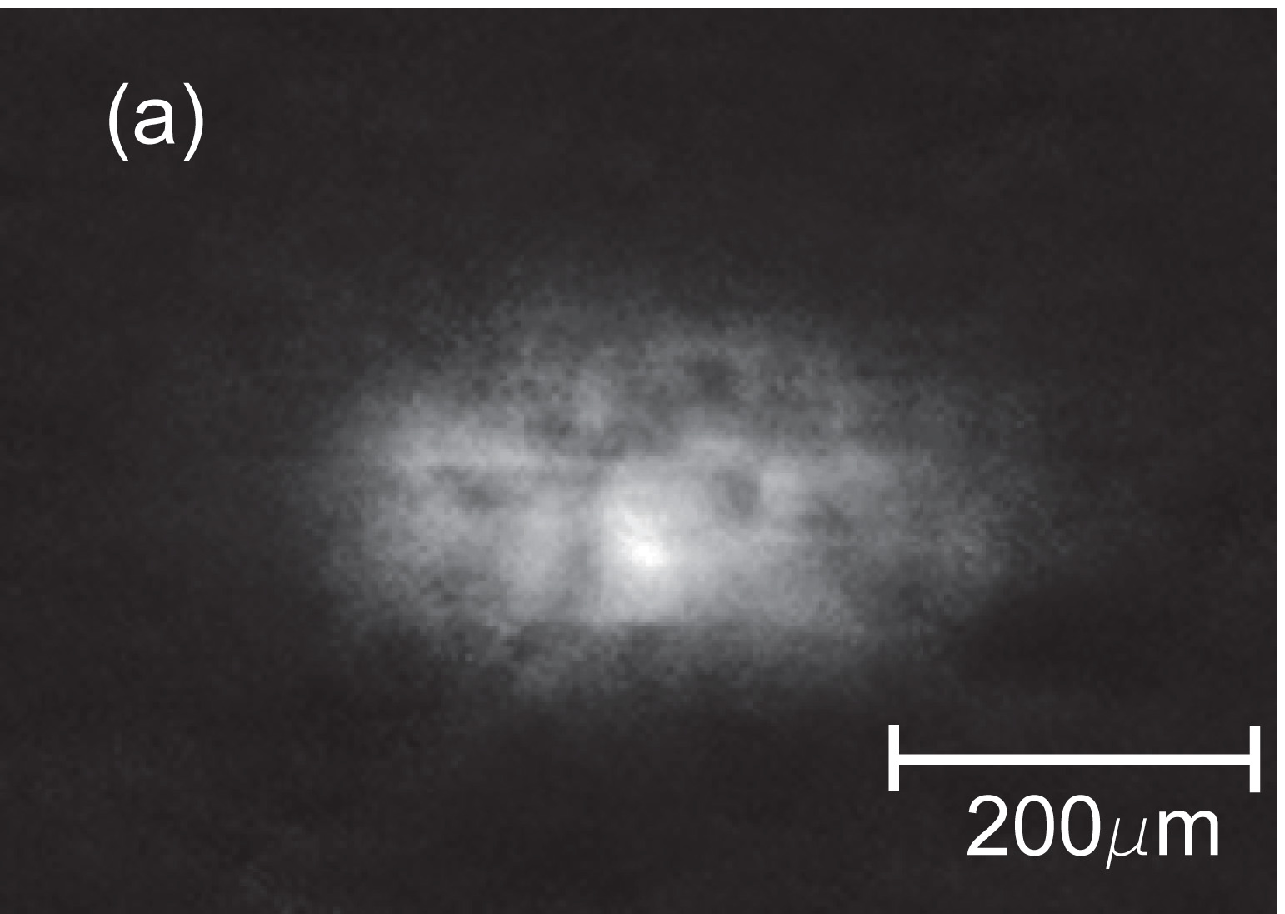}
 \includegraphics[scale=0.35]{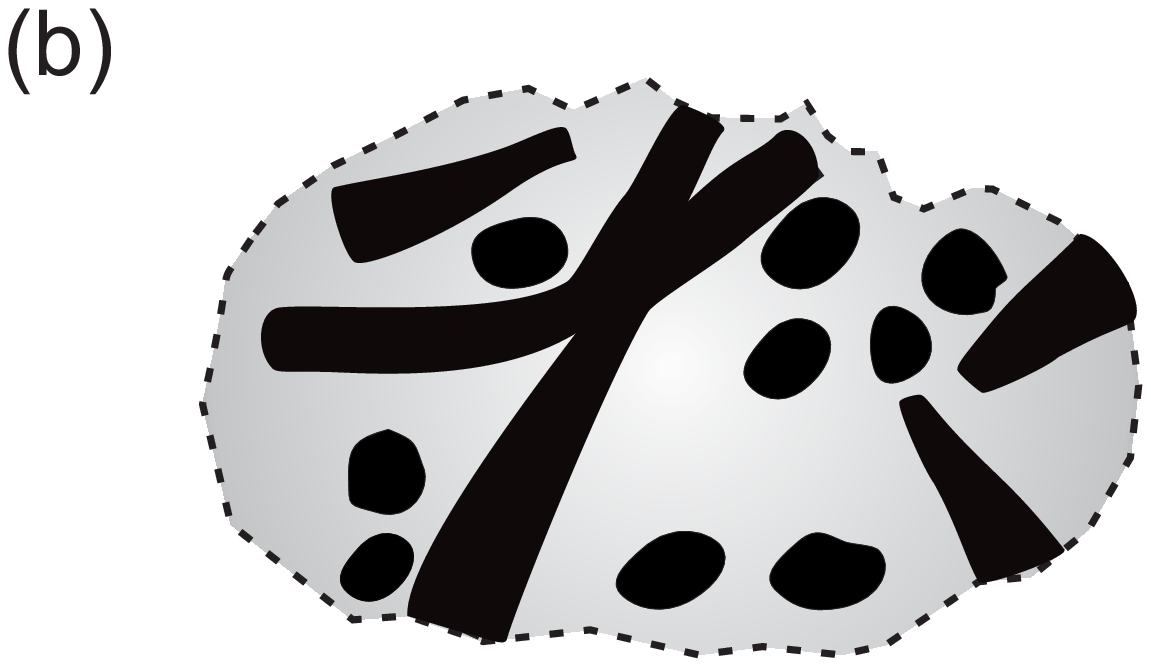}
 \caption{(a) Atomic optical density after 15ms of free expansion showing vortex structures spread all around the cloud resembling the vortex tangle regime proposed in Ref.\cite{tsubotaturb} and (b) Schematic diagram showing the inferred distribution of vortices as obtained from image shown in (a).}
\label{turb}
\end{figure}

When the amplitude and/or excitation time is increased above some given values a completely different regime takes place. We observe a fast increase in the number of vortices followed by a proliferation of vortex lines not only in the original plane of the individual vortices, but distributed for the whole sample, covering many directions. An example of such a regime can be seen in Fig.\ref{turb}a. In fact, equivalently to the already reported instability in the vortex formation \cite{vortexform}, here we also observe an evolution from distinct vortices to the turbulent regime, where many vortices without preferred orientation are formed. Following the nomenclature of Tsubota \textit{et al.}\cite{tsubotaturb}, we call this regime a vortices tangle configuration. Besides the fact that the image in fig.\ref{turb}a comes from absorption in a single direction, careful observation allows one to identify lines crossing the sample in many angles. For the purpose of visualization, together with the real absorption image we plot lines that indicate the presence and position of the vortices in the turbulent cloud in Fig.\ref{turb}b.

The presence of quantized vortices non regularly distributed along the whole sample in all directions is quite characteristic of turbulence in the quantum fluid and it is here taken as evidence for this regime. In fact, our way of nucleating vortices, where non-equivalent translational and torsional movements of the quantum cloud occur in different planes of symmetry of the sample, is closely related to what is proposed in \cite{tsubotaturb} for combined rotations. 

An important question relies on which is the maximum number of vortices that can be introduced in a BEC in the turbulent regime. This number is certainly much smaller than in liquid Helium, mainly restricted by the available fluid volume (smaller in the BEC) and vortex core size (bigger in the BEC). Consequently some turbulence features might not be observable and/or be very different compared from that encountered in other cases of quantum turbulence, which involve large volumes containing many vortices, specially those related to the Komolgorov spectra, where large volumes are needed. Nevertheless, that low vortex density may allow one to make detailed observation of vortex reconnections and all the related phenomena such as cascade-like processes.

Since the absorption imaging is destructive, in the present experiment it is not possible to observe the evolution of turbulence, vortex reconnections, Kelvin-wave cascades and all the dynamic behavior related to turbulence. Nevertheless, with the in-situ, non-destructive imaging technique of phase-contrast \cite{phase}, it will be certainly possible to observe the whole dynamics of the turbulent regime in cold atoms, from the transition from the regular vortex pattern to the turbulent pattern, going through vortex reconnections ending in the mechanisms that eventually lead to the decay of turbulence. Besides, imaging the atoms with this technique from two or more different directions, it is even possible to reconstruct \cite{reconstruct} the whole tridimensional vortex tangle distribution. 

Focusing on the turbulent regime, although we cannot predict the number of vortices formed nor their relative positions/orientations, we do not rely only on the visual signature of the cloud with plenty of vortices to assign the presence of turbulence. In fact, another remarkable feature is present in this cloud to confirm that it follows a completely different hydrodynamic regime: the suppression of the aspect ratio inversion during free expansion. This phenomena is only present after the turbulent regime takes place. The cloud does invert its aspect ratio when only a few regular vortices are present.

As it is widely known, one of the most strong signatures of the quantum degeneracy in a gas of bosons is its asymmetric free expansion \cite{becexp}. In contrast to the isotropic expansion of a thermal cloud, an initially anisotropic trapped condensed cloud expands faster in the most confined directions than in the less confined, so a cigar-shaped cloud, when free to expand, turns rapidly into a pancake-shaped cloud with an inversion of its aspect ratio. This phenomena is well explained theoretically \cite{castin} and is basically due to the domain of the interatomic interaction energy over the kinetic energy in the condensate. The former is proportional to the atomic density distribution which is anisotropic.

In contrast to either the thermal and superfluid free expansion behaviors, the turbulent cloud keeps its aspect ratio during the whole free expansion or, in other words, the aspect ratio inversion is suppressed. Still more remarkable is the fact that it does not evolve to the unitary aspect ratio as in a thermal gas, evidencing that this cloud is certainly not thermal, keeping somehow a quantum nature as it should be expected in a true quantum turbulent regime. 

\begin{figure}
\centering
\includegraphics[scale=0.4]{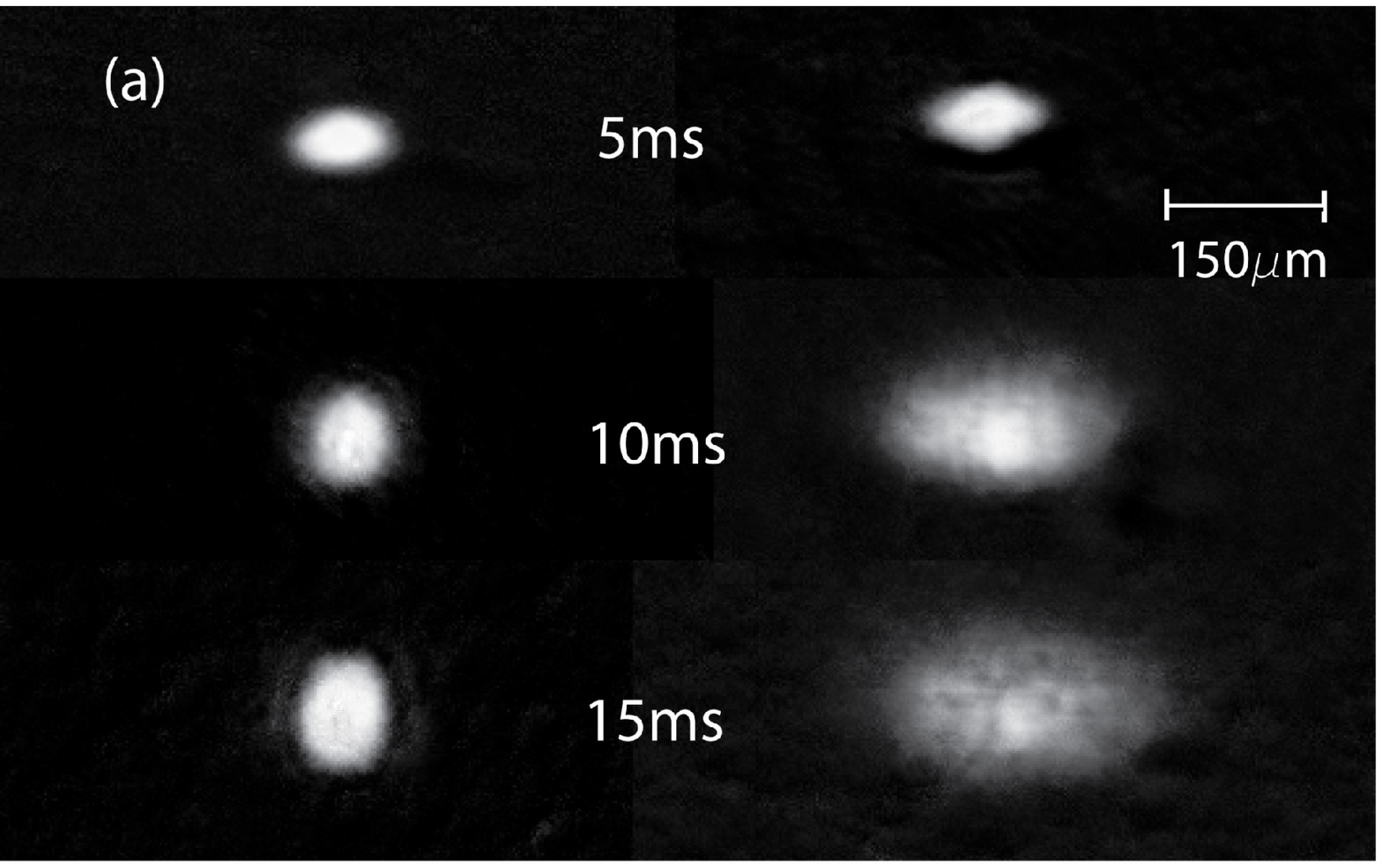}
\includegraphics[scale=0.28]{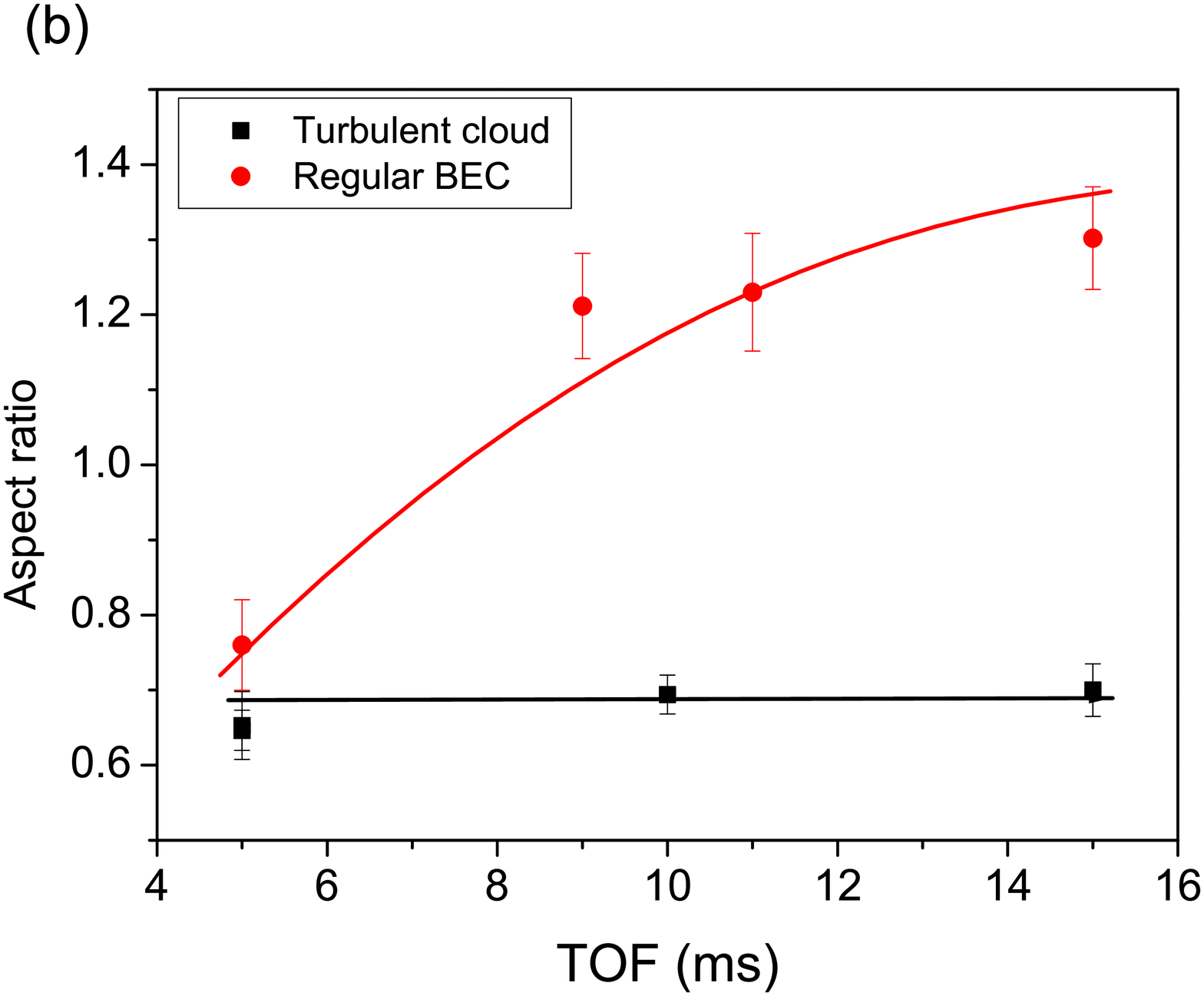}
\caption{(a) Comparison between the free expansion of a normal BEC showing the well known aspect ratio inversion (left) and a turbulent cloud (right) that keeps its aspect ratio while expanding. (b) Quantitative plot of the aspect ratios of the normal BEC cloud (circles) and the quantum turbulent cloud (squares). The lines are only a guide for the eye to stress the observed behavior.}
\label{ar}
\end{figure}

Figure \ref{ar}a shows side by side several normal BEC clouds and turbulent clouds at different times of free expansion, showing the described behavior. Fig.\ref{ar}b shows the quantitative aspect ratio for the turbulent regime (black squares) and for the normal BEC (red circles). We believe that this remarkable characteristic is a new effect for atomic superfluids, possibly as a signature of the emergence of QT in this system.

The change in the aspect ratio during free expansion of a vortex populated quantum fluid is a topic still not well explored in the literature. On the other hand, studies of hydrodynamic expansion of a rotating vortex free superfluid cloud have already been presented \cite{rotbec}. In that case, it was observed that conservation of angular momentum combined to the irrotational flow, causes the expansion to take place in a distinctively way when compared to the non rotating superfluid. In our case, in contrast, the expanding cloud is populated of many vortices in many directions and just keeps its aspect. We believe that the presence of this random flow field along the condensate somehow acts preventing the known expansion regimes. A full theoretical understanding of this topic is still missing and it would certainly shine some light on the true nature played by the flow of matter around the vortex cores and their interplay with interatomic interactions, with consequences on the hydrodynamics of the quantum fluid. Also, a full theoretical understand of this matter might help in the understanding of turbulence itself. A comparison of the expanding BEC and turbulent BEC (Fig.\ref{ar}a) shows that the latter is typically much bigger for longer expanding times, which may be related to the decay of the turbulence through vortex line reconnection and excitation of Kelvin-waves and possible existence of cascades phenomena \cite{cascades}, leading to a heating of the sample. 

In conclusion we have shown experimentally the emergence of quantum turbulence in an atomic quantum fluid evidenced by the presence of vortex cores as well as vortex lines in the absorption imaging non regularly distributed across the sample. These configurations are known as vortex tangles, in contrast to the crystallization regimes, and are an important characteristic of the turbulent regime. Also, we have shown that the turbulent cloud does not expand like either a quantum or a thermal gas, but keeps a fixed aspect ratio. This expansion regime has never been observed and is also believed to be a signature of the presence of QT. With the assumption that the absence of inversion is a feature indicating turbulence in the quantum fluid it could also be used to indirect quantify this transition as well as being an indicative of a completely new hydrodynamic regime of the atomic superfluid.

Experiments to observe QT in BEC are being carried out also at NIST, in the group of K. Helmerson \cite{Kris}. We acknowledge helpful suggestions from J. Dalibard, M. Modugno, W. Philipps and C. Salomon and technical support of R.F. Shiozaki.

This work was supported by FAPESP, CNPq and CAPES.

\end{document}